# Optimization of the Repumping Parameters for a Sodium Laser Guide Star Magnetometer

Yucheng Yang[1, 2, a*], Chunyang Lei[3, b], Kai Guo[4,c], and Shuai Wang[5, d]

[1]State Key Laboratory of Spatial Datum, Beijing 100020, China

[2]State Key Laboratory of Photonics and Communications, School of Electronics, and Center for Quantum Information Technology, Peking University, Beijing 100871, China

[3]AthenaEyes Co., Ltd., Changsha 410205, China

[4]Institute of Systems Engineering, AMS, Beijing 100141, China

[5]Institute of Optics and Electronics, CAS, Chengdu 610209, China

[a]yangyucheng@pku.edu.cn, [b]lcy98311@gmail.com, [c]guokai07203@hotmail.com, [d]wangshuai@ioe.ac.cn

**Abstract.** A sodium laser guide star operated as a mesospheric magnetometer modulates a 589 nm laser at the local Larmor frequency and usually diverts a fraction $q$ of its power to a repumping light that recovers atoms lost to the dark $F$ = 1 ground state. The repump light is therefore a set of parameters to be optimized rather than a monotone gain. We map the return flux and the ground-state spin polarization of the full sodium D2 Zeeman structure with density-matrix simulations of the first five modulation periods, over two parameter planes—repump fraction against peak irradiance and duty cycle. The flux-optimal fraction obeys a threshold law in the mean irradiance $\bar{I} = d_c I_p$, $q_{opt} = 0.20\ (1 - 3.2\ (\mathrm{W/m^2})/\bar{I})$ with $R^2 = 0.99$. The polarization, read out for the most strongly driven velocity group, calls for 2.8 times the flux-optimal fraction, and a shot-noise figure of merit combining the two observables for 2.0 times—beyond the range commercial guide star lasers provide. These results give an applicable rule for allocating repump power in a sodium guide star magnetometer and mark the ground-state polarization.

## 1. Introduction

Sodium laser guide stars (LGSs), produced by exciting the mesospheric sodium layer at 85–100 km with a ground-based 589 nm laser [1-4], also support remote atomic magnetometry [5-7]: because the stimulated radiation driven by the resonant circularly polarized pumping light depends on the atomic spin polarization, which precesses about the geomagnetic field at the Larmor frequency, intensity modulation at the Larmor frequency drives a magneto-optical resonance in the return flux whose position gives the field magnitude at the sodium layer. The scheme has been demonstrated with amplitude modulation [8,9], polarization modulation [10], and gated photon counting [11], and characterized in cell experiments [12,13], and it is the only ground-based access to the geomagnetic field at those altitudes.

Every such system contends with the sodium ground-state hyperfine structure. Circularly polarized D2a light ($3S_{1/2}$, $F = 2 \rightarrow 3P_{3/2}$) pumps atoms toward the cycling transition, but spontaneous decay leaks population into the $F$ = 1 manifold, 1.772 GHz below and dark to the D2a drive [14,15], so a repumping component on the D2b line (here +1.713 GHz from the pumping) returns those atoms to $F$ = 2 [15,16]. In a guide star that component carries a fraction $q$ of the same laser power, withheld from the D2a drive, which makes repumping an optimization; continuous-wave (CW) modeling for astronomy puts the optimum at $q$ ~ 0.1–0.2 [15-17], and commercial guide star lasers offer a repumping adjustable up to about 15% of the total power [18]. Magnetometric operation changes the terms of the compromise: Larmor precession about an oblique field destroys the orientation built by steady pumping [19], so the light is gated into an on-window of duration $d_c T_L$ per Larmor period $T_L$ [6,20,21], with duty cycles $d_c$ = 0.1–0.3 in the demonstrated systems [9,13]. Two observables matter rather than one—the return flux, which sets the photon-shot-noise floor of the magnetometer, and the ground-state spin polarization, which sets the magneto-optical resonance contrast and hence the field-measurement precision. Existing pulse-format optimizations address formats that are not Larmor-synchronous and take the flux as the sole figure of merit [22,23]. For this regime neither the size of the repumping optimum nor the quantity that controls it has been reported, and the practical question is whether the requirement is set by the repump power $qI_p$—a small fraction of a strong beam suffices—or by the fraction $q$ itself, in which case the split ratio must be engineered. The two rules give opposite advice when the laser power changes.

In this paper, we focus on the optimization of the repump light for a guide star magnetometer—by mapping both observables over two parameter planes with density-matrix simulations of the full sodium D2 Zeeman structure: repump fraction against peak irradiance and against duty cycle. Three results emerge: neither candidate rule holds; the flux-optimal fraction follows a threshold law in the mean irradiance $\bar{I} = d_c I_p$, which identifies the cycle-averaged scattering rate as the controlling quantity; the polarization requires 2.8 times as much repumping as the flux does, beyond the range commercial systems provide. Section 2 states the model, the observables, and—explicitly—the simulation protocol and its measured limitations; Section 3 presents the maps, the threshold law, and the design implications; Section 4 states the conclusions.

## 2. Theoretical Model and Simulation Conditions

The atomic evolution is computed from the optical Bloch equations of the full D2 system—the 24 Zeeman sublevels of $3S_{1/2}$ and $3P_{3/2}$—including optical excitation, spontaneous decay, Larmor precession, spin-damping and velocity-changing collisions, and beam-volume exchange, using the open-source AtomicDensityMatrix/LGSBloch Mathematica packages [16,24] that underlie previous LGS magnetometry studies [9,17,25]. Doppler broadening is treated by discretizing the velocity distribution along the beam into velocity groups, and photon recoil by transferring excited-state population to the adjacent blue-shifted group on decay [17]. The laser intensity is square-wave modulated at the Larmor frequency $f_L = \gamma B$, on for the first $d_c T_L$ of each period, and the center frequency is chirped linearly at a rate α applied continuously across on- and off-windows. Table 1 lists the fixed conditions. The geomagnetic and geometric values correspond to the sodium LGS testbed at the Changping campus of Peking University, and the temperature and collision rates follow previous LGSBloch studies [9,17]. The linewidth of 5 MHz is the standard single-frequency output of commercial 589 nm systems [18], a choice whose effect on guide star brightness has been modelled separately [26], the duty cycle of 0.2 lies in the demonstrated range and is swept in one scan, and the chirp rate of 0.5 MHz/μs is held fixed throughout.

Table 1. Fixed simulation conditions (unless swept).

|  |  |
|---|---|
| Transition | Na D2, 589.159 nm, circular polarization |
| Repump (D2b) detuning | +1.713 GHz; power fraction $q$ (swept) |
| Laser lineshape | Lorentzian, FWHM 5 MHz |
| Geomagnetic field $B$; zenith/azimuth | 0.55062 G; 121.64° / −7.32° |
| Larmor frequency $f_L$; period $T_L$ | 385.3 kHz; 2.595 μs |
| Duty cycle $d_c$ | 0.2 (swept in one scan) |
| Peak irradiance $I_p$ | 50 W/m$^2$ (swept in one scan) |
| Chirp rate α | 0.5 MHz/μs |
| Temperature | 185 K |
| Spin-damping / velocity-changing collisions | 1/(245 μs) / 1/(35 μs) |
| Recoil shift per scattering event | 50 kHz |
| Periods simulated per grid point | 5 (from the dark initial state) |

Two planes were scanned, each over $q$ = 0–0.5 in steps of 0.01: against peak irradiance $I_p$ = 1–101 W/m$^2$ in steps of 2.5 (2091 points, $d_c$ = 0.2); against duty cycle $d_c$ = 0.05–0.7 in steps of 0.025 (1377 points, $I_p$ = 50 W/m$^2$). The scans share parameter lines and reproduce each other on them to the last digit, as deterministic simulations of one model must. The study reports per-atom specific return at prescribed mesospheric irradiance: atmospheric propagation, launch optics, and spot averaging are excluded common factors, and the repump light is modeled as a spectrally identical component of the same laser, which is precisely what makes the power split a constraint—systems with an independent D2b source pay a different price.

The specific return flux Ψ (photons $s^{-1}$ $sr^{-1}$ $atom^{-1}$ / ($W/m^2$)) is the cycle-averaged backward fluorescence per atom normalized to the mean total irradiance $\bar{I} = d_c I_p$, so that a change of $q$ redistributes power without changing $\bar{I}$ and comparisons at fixed $\bar{I}$ are comparisons at equal launched power. The ground-state spin polarization is defined as $P = \Sigma_m\, m\, \rho(m)$ [27], where $m$ is the magnetic quantum number of a Zeeman sublevel and $\rho(m)$ its population. We evaluate it from the $F = 2$ populations $P$ of the single most strongly driven velocity group (selected at initialization), normalized within that group, at the end of the fifth on-window; it quantifies the orientation produced in the velocity class the laser addresses, which governs the magneto-optical resonance contrast [6,9].

Both observables are read at the fifth modulation period. The window (13 μs) is short compared with the relaxation times in the model (velocity-changing collisions ≈ 13 periods, spin damping ≈ 94 periods), so the maps report the early transient response of the pumping process rather than a long-time limit. That is a deliberate feature of the protocol, and the reason is structural: under the continuously chirped excitation used throughout, the laser detuning advances monotonically for as long as the run continues, so the drive never repeats and no stationary regime exists for the simulation to converge to. Long runs at selected points confirm this directly: the per-period flux rises, turns over, and then declines monotonically to the end of every run, by 4–19% per fifty periods, with no plateau anywhere within 200 periods. A longer window would therefore not be a better-converged measurement, only a later one, and its value would depend on where the run was stopped. A short fixed early window removes that arbitrariness: every point of both planes starts from the same unpolarized state and is sampled after the same elapsed time under the same drive history, so the comparison across a plane is controlled even though the absolute level is transient. The maps are read in that spirit — as a controlled comparison of how the repump fraction redistributes population in the atomic ensemble, not as a prediction of absolute photon return. Two assumptions follow and are carried through the paper. First, every value quoted below is a five-period transient value under this one protocol, so comparisons and the location of optima are internally consistent, while the numerical constants of Section 3.2 are constants of the protocol and are not steady-state values. Second, the polarization is read out for the most strongly driven velocity group and therefore exceeds the velocity-integrated ensemble value, which at five periods has itself developed only a small fraction of its eventual magnitude; and where the discrete velocity-group layout re-allocates along a scan, $P$ shifts by steps of order 0.1 while the flux does not move, which confines quotable polarization values to the smooth parts of the computed surface (in the repump–irradiance scan, $I_p < 56$ $W/m^2$). The polarization conclusions accordingly rest on where the optima lie, not on absolute values.

## 3. Results and Discussion

### 3.1 Two Candidate Rules for the Repump Fraction

Fig. 1 shows both observables over the ($q$, $I_p$) plane. The flux has an interior maximum, Ψ = 302.77, at $q = 0.16$, $I_p = 73.5$ W $m^{-2}$, and the benefit of repumping grows with irradiance—relative to $q = 0$ at the same irradiance, the best fraction gains nothing at 11 $W/m^2$ but 7.0% at 51 and 18.5% at 101 $W/m^2$. The polarization has an interior maximum in $q$ at every irradiance, moving to larger $q$ as the irradiance grows, with its largest smooth-region value $P = 1.151$ at $q = 0.38$, $I_p = 53.5$ $W/m^2$.

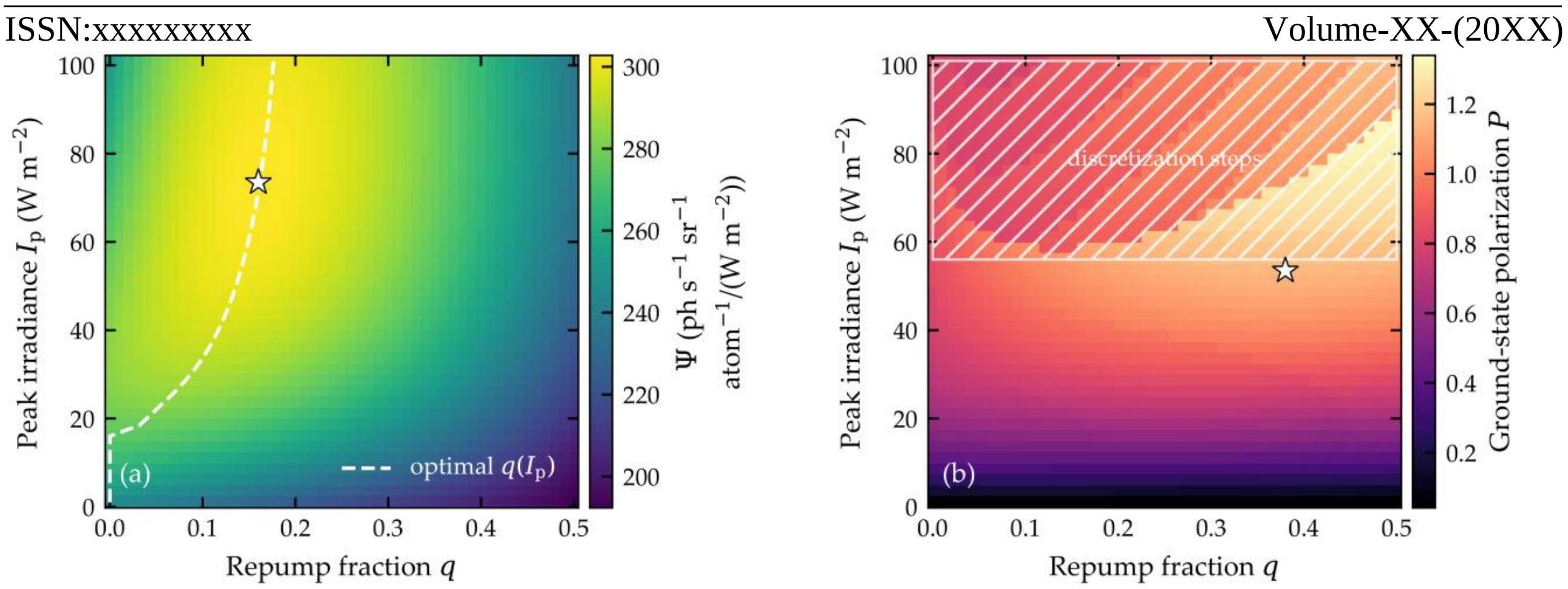

Fig. 1 Repump fraction against peak irradiance at $d_c = 0.2$. (a) Specific return flux $\Psi$; the dashed line traces the flux-optimal fraction and the star marks the maximum at $q = 0.16$, $I_p = 73.5$ W/m$^2$. (b) Ground-state polarization $P$; the hatched band marks the discretization-step region from which no value is read (Section 2), and the star marks the largest smooth-region value, $P = 1.151$ at $q = 0.38$.

The two candidate design rules make opposite, falsifiable predictions about that ridge: a fraction-controlled optimum requires $q_{opt}$ independent of $I_p$, a power-controlled one requires $q_{opt}I_p$ independent of it. Fig. 2 tests both: over $I_p$ = 34–101 W/m$^2$ the optimal fraction runs from 0.107 to 0.176 (a 65% variation) and the optimal repump irradiance from 3.84 to 17.8 W/m$^2$ (363%), so both rules fail, the constant-power rule by the wider margin—and over the full range in which repumping pays at all ($I_p \geq 18.5$ W/m$^2$) the two quantities vary by factors of 5.5 and 30. We quote the narrower range throughout as the conservative statement.

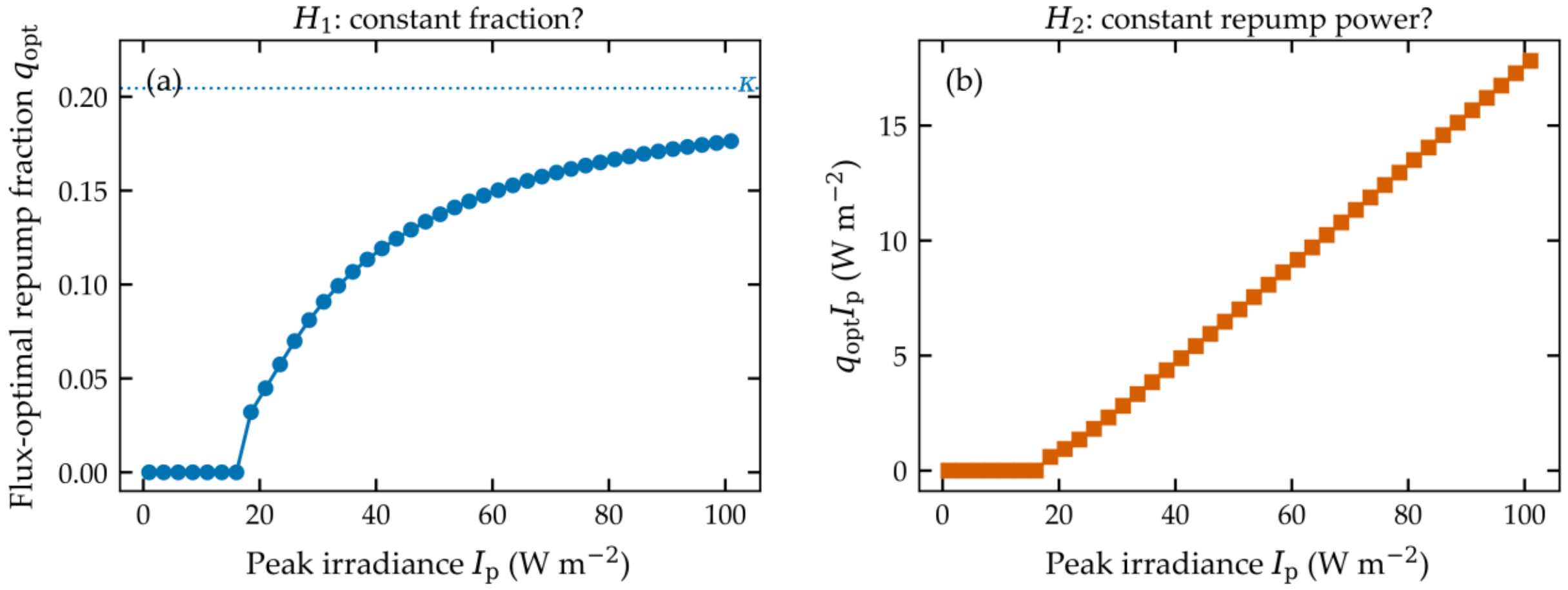

Fig. 2 The two candidate rules, tested on the flux ridge of Fig. 1a. (a) Flux-optimal repump fraction against peak irradiance: zero below $I_p \approx 16$ W/m$^2$, then rising toward the asymptote κ (dotted) of Eq. (1). (b) Optimal repump irradiance $q_{opt}I_p$: nearly linear growth. Neither a constant fraction nor a constant power describes the same ridge.

### 3.2 The Variable That Organizes the Repump Optimum

The variable that does organize the data is the mean irradiance $\bar{I} = d_c I_p$, proportional to the cycle-averaged scattering rate and hence to the loss rate into $F = 1$, whereas the cost of repumping—the power withheld from the drive—is $q$ regardless of $\bar{I}$; balancing a loss that grows with $\bar{I}$ against a cost that does not predicts an optimum that is zero below a threshold and saturates far above it,

$$q_{opt}(\bar{I}) = \kappa(1 - \bar{I}_{th}/\bar{I}) \text{ for } \bar{I} > \bar{I}_{th}; \quad q_{opt} = 0 \text{ for } \bar{I} \leq \bar{I}_{th}. \tag{1}$$

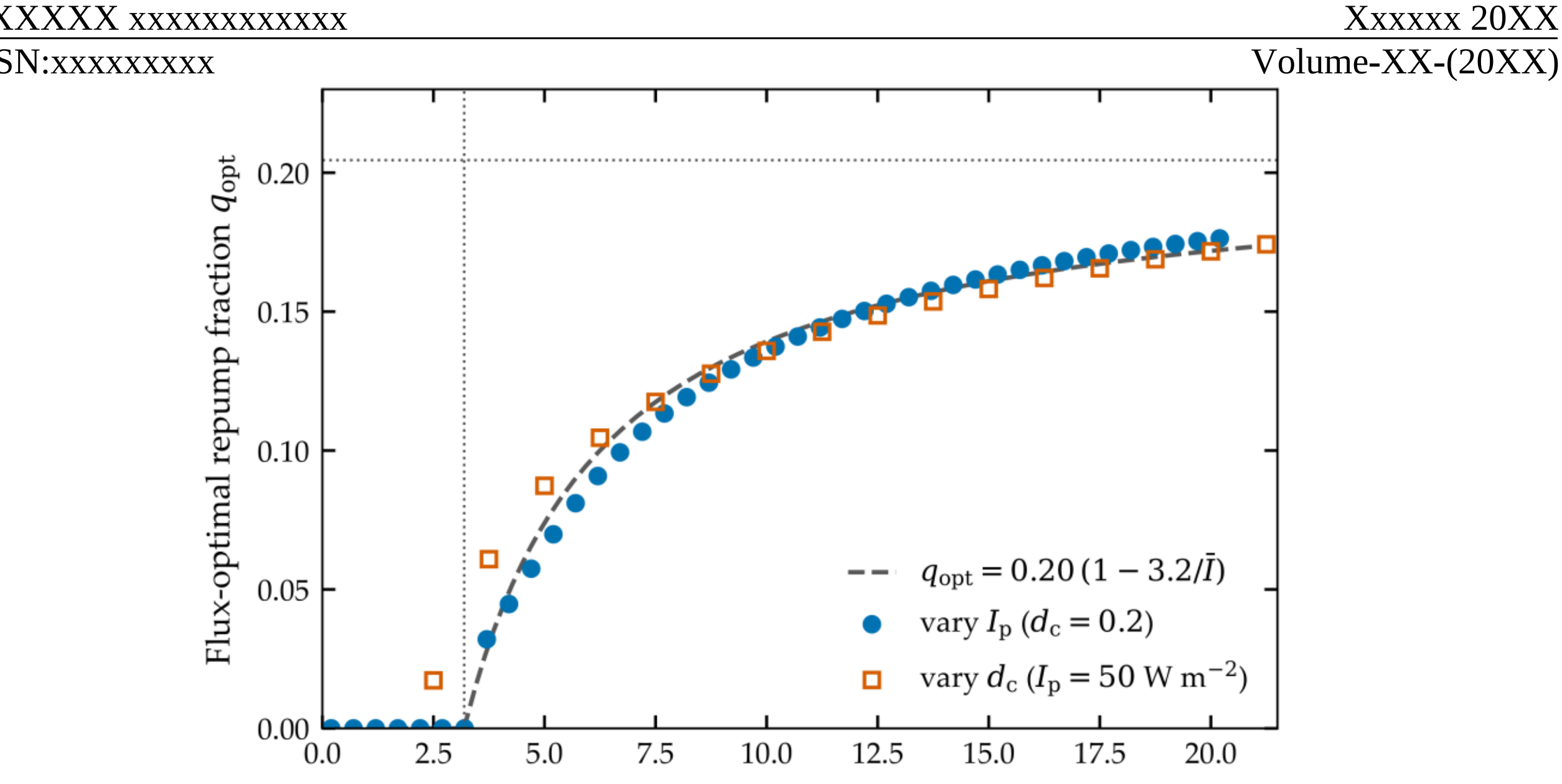

Fig. 3 The flux-optimal repump fraction collapses onto the mean irradiance. Circles: optima from varying $I_p$ at $d_c = 0.2$; open squares: optima from varying $d_c$ at $I_p = 50$ W/m$^2$; dashed: Eq. (1) fitted jointly ($\kappa = 0.20$, $\bar{I}_{th} = 3.2$ W/m$^2$, $R^2 = 0.99$). The two scans reach a given $\bar{I}$ through different variables, so their falling on one curve identifies the controlling quantity.

Fig. 3 tests Eq. (1) against two independent data sets—the optima obtained by varying the peak irradiance at fixed duty cycle and those obtained by varying the duty cycle at fixed peak irradiance—which probe $\bar{I}$ through different factors and collapse onto one curve. A joint fit gives $\kappa = 0.20$ and $\bar{I}_{th} = 3.2$ W/m$^2$ with $R^2 = 0.99$ and an rms residual of 0.006 in $q$ over the 68 points; fitted separately the two scans give $\kappa = 0.208$ and 0.200, $\bar{I}_{th} = 3.37$ and 2.86 W/m$^2$, reproducing the asymptotic fraction to 4% and the threshold to 18%. The collapse carries the conclusion: the mean irradiance—not the peak irradiance or the duty cycle separately—sets the repump requirement.

### 3.3 The Operating Point and the Polarization Requirement

At the operating irradiance $I_p = 51$ W/m$^2$ (Fig. 4) the flux-optimal fraction is $q = 0.137$ with $\Psi = 299.78$; the configuration used in the demonstrated magnetometers [9], $q = 0.12$, gives $\Psi = 299.53$, only 0.08% below the optimum, while removing the repumping light costs 6.5% and raising the fraction to 0.50 costs 23.5%. The polarization is maximized at a much larger fraction: at the same operating point its optimum is $q = 0.38$ with $P = 1.1305$, against 1.0108 at $q = 0.12$ and 0.8334 without repumping—2.8 times the flux optimum, at a price of 11.8% of the flux. That ratio is a property of the single-velocity-group five-period readout. The velocity-integrated ensemble polarization, available from long runs at three fractions on this line, puts its own optimum above the flux optimum under either reading, but at a window-dependent distance: after two hundred periods it rises monotonically across the sampled fractions (0.633, 0.687 and 0.765 at q = 0.12, 0.20 and 0.50), placing its optimum at or beyond 0.50, whereas at five periods, where it has reached only the 3 to 19% of its asymptote quoted in Section 2, the same three values peak at the middle one. The single-group factor 2.8 lies inside the range of roughly 1.5 to 3.6 that those two readings span; where the ensemble optimum falls within that range is left open, since it is a property of the observation window rather than of the repump split. The same fraction reappears in the duty-cycle plane ($P = 1.1219$ at $d_c = 0.20$, $q = 0.38$) smooth at that point and sharing no grid with the irradiance plane apart from one line. The gap has a hardware consequence: the 0–15% repumping range of commercial guide star lasers [18] covers the flux optimum comfortably but falls well short of 0.38, so a contrast-limited magnetometer would need a repumping capability that current systems do not offer.

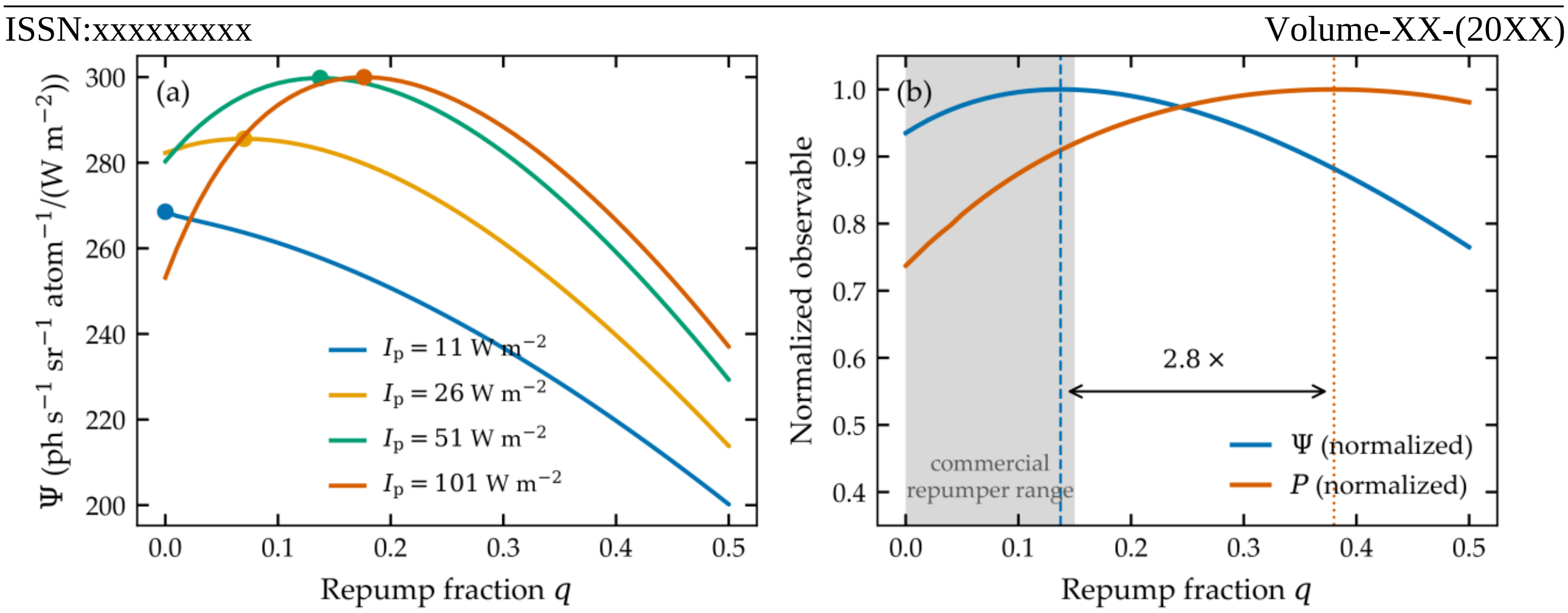

Fig. 4 Repump fraction at the operating point. (a) Flux against $q$ at four peak irradiances; circles mark the maxima, which move to larger $q$ with irradiance. (b) Flux and polarization at $I_p$ = 51 W/m$^2$, each normalized to its own maximum: the flux peaks at $q$ = 0.137 (dashed), the polarization at $q$ = 0.380 (dotted), a factor 2.8 apart. The shaded band is the 0–15% repumping range of commercial guide star lasers [18], which contains the flux optimum but not the polarization optimum.

### 3.4 Robustness of the Two Optima

Two properties of the computed surfaces bound how far the two optima can be carried, and each is measured rather than asserted. The velocity discretization, first: doubling the group count from the default 202 to 402 at fixed run length reproduces the flux to −0.20%, uniformly across a six-point test grid, and the polarization to within 0.002, leaving the step boundary of Fig. 1b in the same place. The steps are therefore not a resolution deficiency. The velocity-group constructor floors the bin width at five recoil shifts, so refinement cannot narrow the bins the resonance actually occupies, and what the steps track is a fixed layout boundary sampled by an instantaneous readout; excluding the step region, as Section 2 does, is the remedy adopted here. Second, the resulting localization: on the smooth part of the surface the flux optimum is pinned to $q$ = 0.137 with an uncertainty of 0.02, the interval lying within the measured 0.20% binning offset of the maximum, whereas $P$ is flat enough about its own maximum that $q$ = 0.38 carries roughly +0.12 and −0.24 at the one-step level. The separation of the two optima, not the second digit of their ratio, is the result that survives this.

The two optima can also be combined instead of chosen between. A magnetometer limited by photon shot noise on a magneto-optical resonance of fixed width has an uncertainty proportional to the reciprocal of the product of the resonance contrast and the square root of the photon rate, so taking the contrast proportional to the ground-state polarization makes P√Ψ the quantity to maximize. At Ip = 51 W/m2 it peaks at q = 0.276, between the flux optimum at 0.137 and the polarization optimum at 0.380, and stays within 1% of its maximum over q = 0.22 to 0.34 (Fig. 5b). The location barely depends on how the two observables are weighted: exponents ranging from PΨ to PΨ^(1/4) move it only across 0.235 to 0.314. Even this compromise fraction is about twice the 0.15 ceiling of commercial repumpers, so the hardware conclusion does not require adopting the polarization optimum outright. Converting the two optima into a sensitivity budget needs the resonance lineshape and the noise spectrum and remains follow-up work; what the combined metric settles here is that no weighting of flux against polarization brings the repump requirement inside the range current lasers provide.

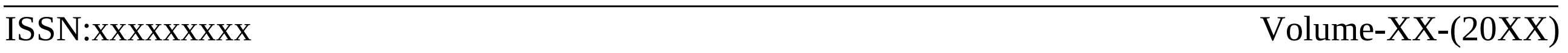

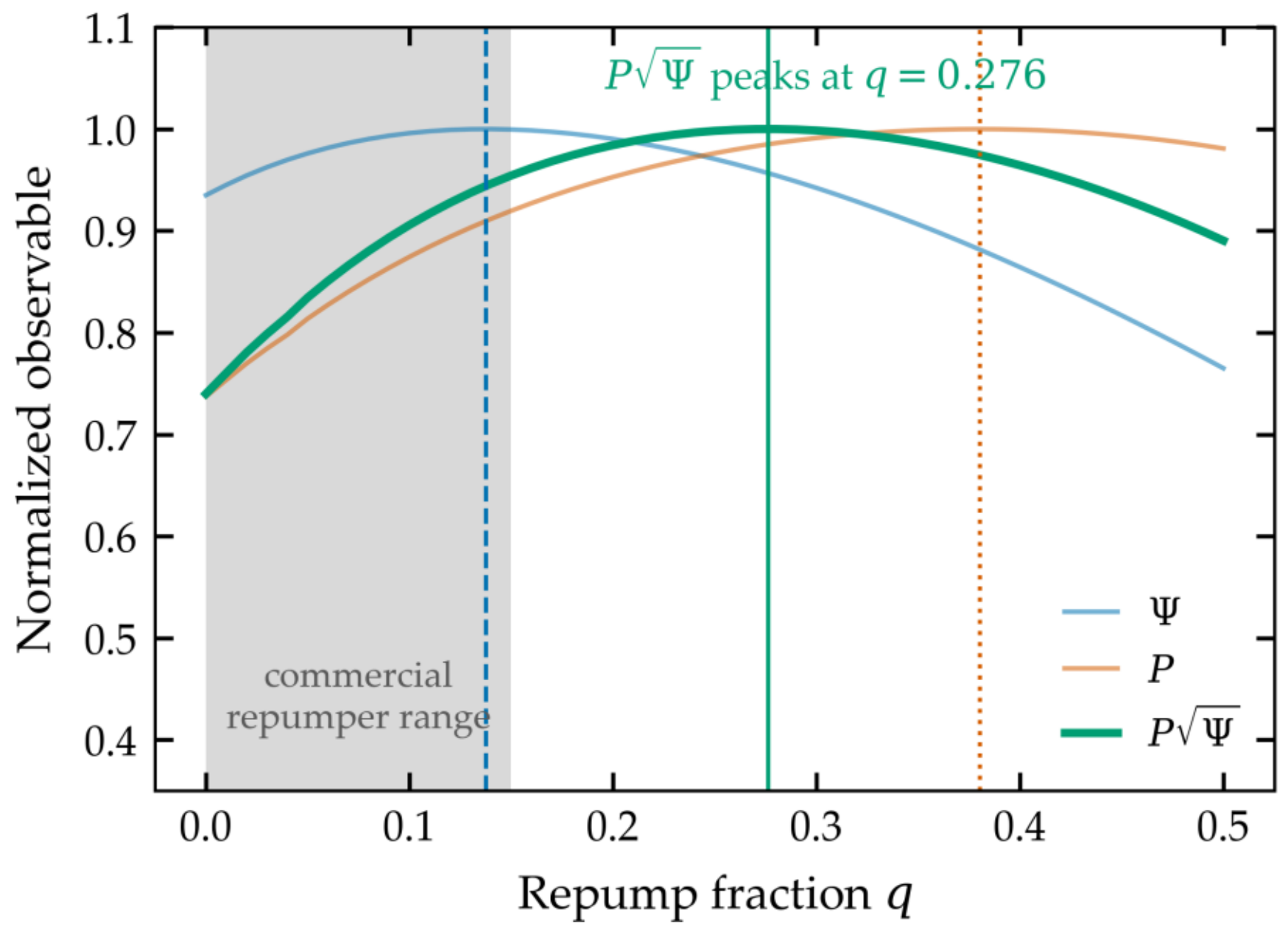

Fig. 5 Robustness of the two optima. Normalized observables at $I_p$ = 51 W/m$^2$ together with the shot-noise figure of merit: its maximum at $q$ = 0.276 falls between the flux optimum (dashed) and the polarization optimum (dotted), and remains above the 0 to 15% commercial repumper range (shaded).

### 3.5 Design Implications and Limitations

The threshold law answers the design question in a form that can be applied without redoing the simulations: the repump requirement is set neither by a fixed split ratio nor by a fixed repump power but by the mean irradiance delivered to the mesosphere, through Eq. (1)—below $\bar{I} \approx 3$ W/m$^2$ the power is better spent on the drive, far above threshold the fraction approaches $\kappa \approx 0.20$ and never exceeds it, and in between it must be scaled with $\bar{I}$ (by 0.03 in $q$ between $\bar{I}$ = 10 and 20 W/m$^2$, for example), so a laser upgrade also changes the repump fraction the guide star requires. The mechanism is a balance between a loss into $F$ = 1 that scales with the scattering rate and a cost (the withheld power fraction) that does not; this argument predicts the form of Eq. (1) but not its constants, whose first-principles derivation from the branching ratios and collision rates is left as follow-up. Read as an instrument-design rule, the two optima bound the achievable sensitivity of the guide star magnetometer from two sides: the flux optimum minimizes the photon-shot-noise floor, while the polarization optimum maximizes the resonance contrast that turns a field-induced frequency shift into a measurable signal, so optimizing the repumping light is optimizing the precision of the remote magnetometer. At the operating point the standing choice $q$ = 0.12 is flux-optimal to within any measurement, and repumping fractions of 0.1–0.2 agree with CW astronomy modeling [15-17] for a different regime, the correspondence being made explicit by the mean irradiance as the shared variable. The flux–polarization gap ($q$ = 0.137 against 0.38) runs opposite to the intuition that a brighter guide star is a better magnetometer, because atoms parked in $F$ = 1 carry no $F$ = 2 orientation, and repumping keeps helping the polarization long after it stops helping the flux; which optimum to adopt depends on whether the sensitivity is photon-shot-noise- or contrast-limited, a trade whose quantification through a resonance-lineshape and noise model is follow-up work, but the polarization-optimal fraction already lies outside what commercial lasers provide [18].

The limitations are stated explicitly rather than left implicit. All values are five-period transient responses under the fixed protocol of Section 2, so $\kappa$ and $\bar{I}_{th}$ are constants of that protocol and should not be transferred to long-engagement systems without recomputation under a longer observation window; the form of the law rests on the rate-balance mechanism, which does not depend on the window. The polarization is a single-velocity-group early-time readout, so its absolute values are not layer polarizations, though the location of its optimum in $q$ — the quantity the design conclusion uses — is a fixed-protocol comparison confirmed on both planes. The repumping shares one power budget with the pumping by construction; independent-source systems pay a different price. The collision rates follow earlier modeling, and mesospheric measurements suggesting values higher by factors of 2–6 [9] would raise both κ and the threshold. Finally, the mean irradiance entering Eq. (1) is mesospheric, and must be converted through atmospheric transmission and the beam profile before the law is applied to a launched-power budget.

## 4. Conclusions

We have mapped the return flux and the ground-state polarization of a remote magnetometer based on a sodium laser guide star over the two planes that the repump fraction spans with the peak irradiance and the duty cycle, and we find that the flux-optimal repump fraction is governed by a threshold law in the mean irradiance, $q_{opt} = \kappa(1 - \bar{I}_{th}/\bar{I})$ with $\kappa = 0.20$ and $\bar{I}_{th} = 3.2$ W/m$^2$; at the operating point the standing choice $q = 0.12$ is flux-optimal within any measurement, and the ground-state polarization calls for 2.8 times the flux-optimal fraction—a contrast requirement that lies beyond the reach of present guide star lasers. For a sodium guide star magnetometer these results turn the repump split from a fixed number into a rule tied to the delivered irradiance, and they mark the polarization, not the flux, as the quantity whose optimization the hardware must be extended to reach. Two questions are left open by this study: a first-principles derivation of κ and $\bar{I}_{th}$ from the branching and collision rates, and a resonance-lineshape and noise model that would turn the two optima into a sensitivity budget for the instrument. Converting the protocol constants into steady-state values is not among them: it would require an excitation scheme in which a stationary regime exists, which is outside the scope of this study.

## Acknowledgment

This work was supported by the National Natural Science Foundation of China (No. 62301377). The authors thank the maintainers of the AtomicDensityMatrix and LGSBloch packages.